\documentclass[aps,pre,onecolumn,noshowpacs,superscriptaddress,groupedaddress,preprintnumbers,nofootinbib,showpacs,12pt]{revtex4}
\usepackage{epsfig,dsfont,amssymb,amsmath,amsthm,amsfonts,amsbsy,mathrsfs}
\usepackage{lipsum}
\usepackage{color}
\usepackage{graphicx}
\usepackage{amsmath}
\usepackage{amssymb}
\usepackage{subfigure}
\usepackage{float}
\usepackage{hyperref}
\usepackage{verbatim}
\def\d{{\rm d}}

\renewcommand{\eqref}[1]{\textrm{Eq.}(\ref{#1})}

\newcommand {\be}{\begin{equation}}
\newcommand {\ee}{\end{equation}}

\begin{document}

\title{Nonequilibrium thermodynamics of coupled molecular oscillators: \\ The energy cost and optimal design for synchronization}
\author{Dongliang Zhang}
\affiliation{The State Key Laboratory for Artificial Microstructures and Mesoscopic Physics, School of Physics, Peking University, Beijing 100871, China}
\author{Yuansheng Cao}
\affiliation{Department of Physics, University of California, San Diego, La Jolla, California 92093, USA}
\author{Qi Ouyang}
\affiliation{The State Key Laboratory for Artificial Microstructures and Mesoscopic Physics, School of Physics, Peking University, Beijing 100871, China}
\affiliation{Center for Quantitative Biology and Peking-Tsinghua Center for Life Sciences, AAIC, Peking University, Beijing 100871, China}
\author{Yuhai Tu}
\affiliation{IBM T. J. Watson Research Center, Yorktown Heights, New York 10598, USA}
\email{yuhai@us.ibm.com}

\begin{abstract}
A model of coupled molecular oscillators is proposed to study nonequilibrium thermodynamics of synchronization. We find that synchronization of nonequilibrium oscillators costs energy even when the oscillator-oscillator coupling is conservative. By solving the steady state of the many-body system analytically, we show that the system goes through a nonequilibrium phase transition driven by energy dissipation, and the critical energy dissipation depends on both the frequency and strength of the exchange reactions. Moreover, our study reveals the optimal design for achieving maximum synchronization with a fixed energy budget. We apply our general theory to the Kai system in Cyanobacteria circadian clock and predict a relationship between the KaiC ATPase activity and synchronization of the KaiC hexamers. The theoretical framework can be extended to study thermodynamics of collective behaviors in other extended  nonequilibrium active systems.
\end{abstract}

\maketitle
\clearpage
\section{Introduction}
Synchronization among a population of interacting single oscillators is ubiquitous in nature \cite{pikovsky2003,strogatz2003}, e.g., Josephson junctions \cite{josephson1964}, circadian clocks \cite{WINFREE196715}, physiological rhythms \cite{glass2001}, neurons firing \cite{pazo2014,montbrio2015}, and communication in cell populations \cite{gregor2010,danino2010}. Synchronization dynamics have been well studied by using theoretical models, in particular, the Kuramoto model \cite{kuramoto1975,Kuramoto, acebron2005, pinto2017critical}. However, relatively little is known about synchronization of molecular  oscillators in cellular systems where the underlying mechanism is governed by biochemical reactions with a small number of molecules and large fluctuations.

Recently, several studies were published on understanding the energetics of individual biochemical oscillators (clocks) for maintaining their phase accuracy and sensitivity~\cite{yuansheng,Udo1,Udo2,gingrich2017fundamental,fei2018}.
Here, we investigate whether and how much additional energy is required to drive interaction (coupling) among individual molecular oscillators to achieve their collective behavior, i.e., synchronization. %Even though interactions between individual oscillators are derived from an energy function that depends on their phase difference,
We find that even conservative exchange interactions between individual oscillators, in combination with the phase dynamics of individual oscillators, break detailed balance and thus continuous energy dissipation is needed to drive the oscillator-oscillator coupling contrary to previous thought ~\cite{pinto2017critical,lee2018}. In a general model of coupled molecular clocks, %we solved the steady state many-oscillator probability distribution function {\it analytically}, which allowed us to compute the free energy dissipation rate and the synchronization order parameter {\it exactly} in the infinite oscillators limit.
we show that synchronization is achieved only %through a second order phase transition
when the energy dissipation reaches a critical value that depends on both the strength and frequency of oscillator-oscillator exchange reactions. Our theory further reveals the optimal choice (design) of the exchange reaction frequency and strength that leads to the maximum synchronization with a given energy budget. Finally, we apply our theory to the Kai system in the circadian clock of \textit{S. elongatus} to understand its molecular mechanism for synchronization .

\section{Models and results}
\subsection{A model of coupled molecular clocks: the global and local dissipative cycles }
\begin{figure}\centering
\includegraphics[width=.9\textwidth]{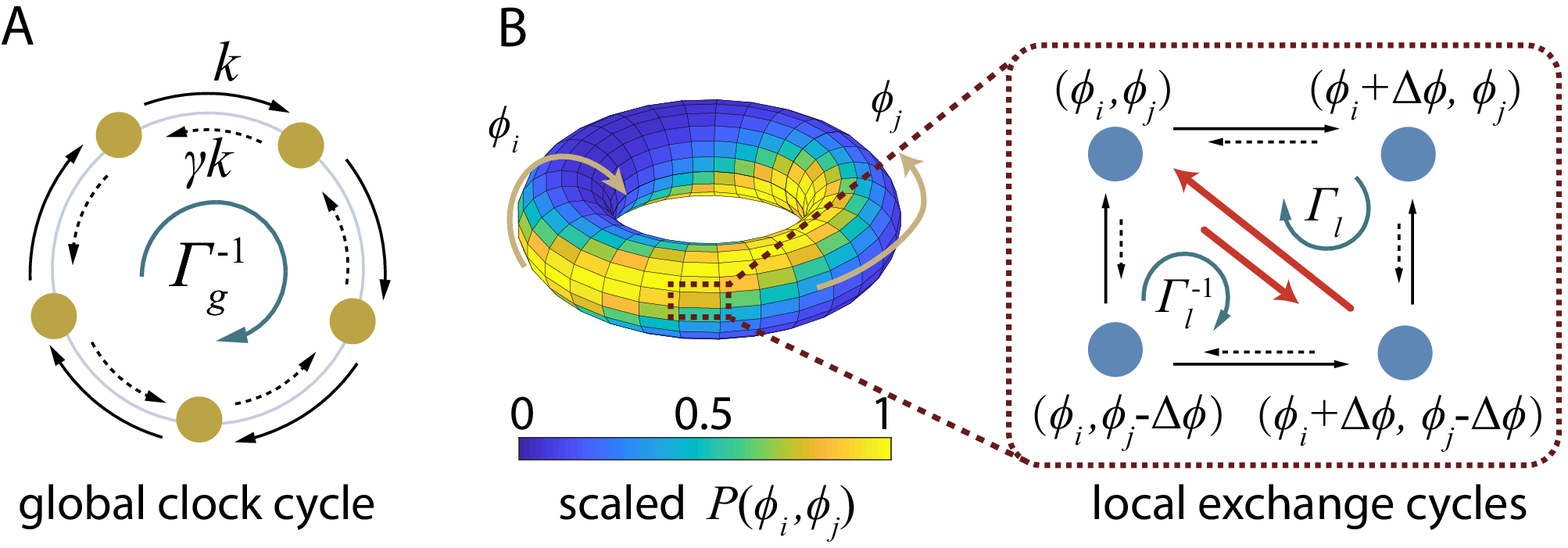}
\caption{Nonequilibrium cycle dynamics of Poisson clock(s). (A) A clock steps between 2 neighboring states by Poisson processes with rates $k$ for clockwise transitions and $\gamma k$ for counterclockwise transitions $(\gamma<1)$. The global clock cycle is characterized by $\Gamma_g=\gamma^{N}$. (B) The distribution function $P(\phi_i,\phi_j)$ of the phases $\phi_i$ and $\phi_j$ of two interacting Poisson clocks are shown on a torus. The transitions among 4 neighboring states in the dotted box are shown with the exchange reactions labeled by red arrows. The two local exchange cycles are characterized by $\Gamma_l(=e^{-\Delta E_{ij}})$ and $\Gamma_l^{-1}$ with $\Delta E_{ij}$ the energy difference between states $(\phi_i,\phi_j)$ and $(\phi_i+\Delta\phi ,\phi_j-\Delta \phi)$.}
\label{coupling}
\end{figure}

%{\bf A model of coupled molecular clocks.}
We consider $m$ interacting molecular clocks, each with $N$ microscopic states labeled by $n=1,2,...,N$. As shown in Fig.~\ref{coupling}A, these microscopic states can be arranged on a ring with a periodic boundary condition, i.e., state $N+1$ is the same as state $1$, and a phase variable $\phi \equiv n \Delta \phi$ is defined. %Dynamics of the clock are determined by chemical reactions between neighboring states in the ring.
In this paper, we study the simple ``Poisson" clock model where both the forward (clockwise) and backward (counterclockwise) transitions between two neighboring states $n$ and $n+1$ are Poisson processes with the forward rate $k_n^+=k$ and the backward rate $k_n^-=\gamma k$. %We call this simple clock model the Poisson clock model.   %The state of the two oscillators are labeled as $(i,j)$, with $i,j\in 1,\cdots,N$. It is easy to check that for the local cycle $(i,j)\rightarrow (i+1,j)\rightarrow (i+1,j-1)\rightarrow (i,j-1)\rightarrow (i,j)$ detailed balance is satisfied. However,

When $\gamma\neq 1$, detailed balance is broken as the products of reaction rates in the counter-clockwise and clockwise directions in the full {\it global} clock cycle $1\rightarrow2\rightarrow\cdots\rightarrow N\rightarrow 1$ become unequal as shown in Fig.~\ref{coupling}A:
\begin{equation} \Gamma_{g}\equiv\prod_n k_n^-/\prod_n k_n^+=\gamma^N\ne 1,
\label{global}
\end{equation}
which means that time reversal symmetry is broken in the system and a sustained oscillation is {\it possible}. Driven by free energy dissipation, reactions along the ring advance the phase of the oscillator \cite{yuansheng,Udo1,Udo2}, and are thus called the processive reactions in this paper.
%The processive stepping of phases can be described by the master equation
%$dP/dt=k P(i-1,j)+\gamma kP(i+1,j)+kP(i,j-1)
%+\gamma kP(i,j+1)-2(1+\gamma)kP(i,j)$ with $P(i,j)$ denoting the probability of the two oscillators at state $(i,j)$.

%{\it Describe large noise and the need for synchrony, and the characterization of interaction ...}
However, spending free energy to keep $\gamma\ne 1$ is only a necessary condition for oscillation in a single clock. Due to large fluctuations in the molecular level chemical reactions (Poisson processes), individual clocks quickly become asynchronous and macroscopic (averaged) oscillatory behavior disappears. To achieve synchronous oscillation, we introduce coupling between two individual clocks $i$ and $j$ as shown in Fig.~\ref{coupling}B (red reaction arrows in the right panel). Specifically, we introduce exchange reactions between the two-clock states $(\phi_i,\phi_j)$ and $(\phi_i+\Delta\phi,\phi_j-\Delta\phi)$, which only change their relative phase but preserve their total phase. The exchange reaction rates are governed by the interaction energy $E(\phi_i-\phi_j)$ that depends on the phase difference of the two clocks:
\[
k_{ex}((\phi_i+\Delta\phi,\phi_j-\Delta\phi)\to (\phi_i,\phi_j))=\frac{\Omega}{m}\exp(-\Delta E_{ij}/2),
\]
\[
k_{ex}((\phi_i,\phi_j)\to (\phi_i+\Delta\phi,\phi_j-\Delta\phi))=\frac{\Omega}{m}\exp( \Delta E_{ij}/2),
\]
%and $k_{ex}((\phi_i,\phi_j) \rightarrow (\phi_i+\Delta\phi,\phi_j-\Delta\phi))=\frac{\Omega}{m}\exp(-\Delta E_{ij}/2)$,
%\begin{eqnarray}
%k_{ex}((\phi_i+\Delta\phi,\phi_j-\Delta\phi) \rightarrow (\phi_i,\phi_j))&=&\frac{\Omega}{m}e^{-\frac{\Delta E_{ij}}{2}}, \\
%k_{ex}((\phi_i,\phi_j) \rightarrow (\phi_i+\Delta\phi,\phi_j-\Delta\phi))&=&\frac{\Omega}{m}e^{\frac{\Delta E_{ij}}{2}},
%\end{eqnarray}
where $\Delta E_{ij}=E(\phi_i-\phi_j)-E(\phi_i-\phi_j+2\Delta\phi)$ and $\Omega$ is the average exchange frequency per oscillator. Other choices of the exchange reaction rates do not change the results in this study (see SI for details).%The factor $m^{-1}$ is in the sense that the overall frequency of each individual oscillator tries to react with others should keep the same and thus the frequency of each reaction should be scaled by the oscillator number.

Note that the ratio of the forward and backward exchange reaction rate is equal to $e^{-\Delta E_{ij}}$, the same as in an equilibrium system with energy function $E(\phi_i-\phi_j)$ and the thermal energy $k_B T=1$. However, these seemingly conservative exchange interactions cost energy in the final nonequilibrium steady state (NESS). This additional energy cost has an intuitive origin as we take a close look at the triangular {\it local} exchange cycle formed by the combination of  two processive reactions and one exchange reaction: $(\phi_i,\phi_j)\rightarrow (\phi_i+\Delta\phi,\phi_j) \rightarrow (\phi_i+\Delta\phi,\phi_j-\Delta\phi)\rightarrow (\phi_i,\phi_j)$ as shown in Fig.~\ref{coupling}B. It is easy to show the ratio of the products of the reaction rates in the clockwise and counter-clockwise directions for this local cycle is:
\begin{equation}
\Gamma_{l}= e^{-\Delta E_{ij}}\times \gamma^{-1}\times \gamma =  e^{-\Delta E_{ij}} \ne 1,
\end{equation}
or $\Gamma_l^{-1}$ for the accompanying local cycle: $(\phi_i,\phi_j)\rightarrow (\phi_i+\Delta\phi,\phi_j-\Delta\phi) \rightarrow (\phi_i,\phi_j-\Delta\phi)\rightarrow (\phi_i,\phi_j)$. The existence of this dipole of cycles ($\Gamma_l\ne 1$)
indicates the violation of detailed balance at the local level in addition to the global violation due to full period phase procession (Eq.~\ref{global}). Therefore, additional energy must be dissipated to drive the exchange reactions for synchronization.

%The exchange reactions should only affect the phase difference between two oscillators but keep their average phase unchanged. A simple scheme of exchange reactions would be:
%$f(i,j)=\text{rate}[(i,j)\rightarrow(i+1,j-1)]=\alpha\exp(-\Delta E/2)$ with its reverse reaction
%$b(i+1,j-1)=\text{rate}[(i+1,j-1)\rightarrow(i,j)]=\alpha\exp(\Delta E/2)$,
%where $\Delta E=E(i+1,j-1)-E(i,j)$, and the $\alpha$ is the exchange rate. The master equation will be the processive stepping (introduced above) added with terms $f(i-1,j+1)P(i-1,j+1)+b(i+1,j-1)P(i+1,j-1)-[f(i,j)+b(i,j)]P(i,j)$.

\subsection{An analytical solution for the many-oscillator phase distribution}
In the limit $N\rightarrow\infty$, the phase of each oscillator can be described by a continuous phase variable $\phi _i\equiv n_i \Delta\phi$. By rescaling reaction rates with $\Delta\phi$ accordingly:  $k(\Delta\phi)^2\to k$, $\Omega(\Delta\phi)^2\to\Omega,$ we obtain the Fokker-Planck equation for the joint distribution function of all the oscillator phases $P(\phi _1,\phi_2,...,\phi_m,t)$ :
\begin{equation}\label{FP_m}
\frac{\partial P}{\partial t}=k\sum_i\frac{\partial}{\partial\phi_i}\left(-e_g+\frac{\partial}{\partial\phi_i}\right)P
+\frac{\Omega}{m}\sum_{i< j}\frac{\partial}{\partial\varphi_{ij}}\left(2E^\prime(\varphi_{ij})+\frac{\partial}{\partial\varphi_{ij}}\right) P,
\end{equation}
%\be
%\frac{\partial P}{\partial t}=k\sum_i\frac{\partial}{\partial\phi_i}(-e_gP+\frac{\partial P}{\partial\phi_i})+2\alpha\frac{\partial}{\partial\varphi}(2E^{\prime}(\varphi)P+2\frac{\partial P}{\partial\varphi}),
%\label{FP_2}
%\ee
where %the subscript $(i\ne j)$ is summed over all oscillator pairs and the exchange frequency $\Omega$ is normalized by the number of oscillators.  T
$\varphi_{ij}=\phi_i-\phi_j$  is the relative phase variable and $\partial/\partial\varphi_{ij}=\partial/\partial\phi_i-\partial/\partial\phi_j$. In the continuous limit, the net speed of phase procession is $k e_g$ with  $e_g=\lim_{N\rightarrow \infty} \ln(\gamma^{-1})/\Delta\phi=-\ln\Gamma_{g}/2\pi$.

The physical meaning of the Fokker-Planck equation, Eq.~\ref{FP_m}, is clear. The first term on the right hand side (RHS) is due to the processive reactions of individual clocks, while the 2nd term on the RHS is due to the clock-clock interaction. Remarkably, the steady state distribution of the coupled many-oscillator system can be obtained analytically with a simple solution (see Methods for derivation):
\begin{equation}
P_s(\vec{\phi})=Z^{-1}\exp(-\beta E_t(\vec{\phi})),
\label{ansatz}
\end{equation}
where $E_t=\frac2m\sum_{(i< j)}E(\phi_i-\phi_j)$ is the total exchange interaction energy, $Z$ is the normalization constant (or the partition function), and the effective inverse temperature parameter $\beta$ equals:
%\begin{equation}
$\beta=\frac{\Omega}{\Omega +k}$.
%\end{equation}

It is important to point out that even though the steady state phase distribution given in Eq.~\ref{ansatz} follows a Boltzmann distribution, the system is in a nonequilibrium steady state (NESS) with an effective nonequilibrium  temperature:
\begin{equation}
T_{eff}\equiv \beta^{-1}=1+k/\Omega, 
\end{equation}
which is higher than the thermal equilibrium temperature (set to unity in our study). The nonequilibrium processive reactions increase the effective temperature by $k/\Omega$ without changing the exchange interaction energy $E_t$. %Furthermore, the average phase speed $d\langle\phi_i\rangle_i /dt=ke_g$  is independent of the coupling.

From the steady state distribution $P_s$ given by Eq.(\ref{ansatz}), we can compute the probability flux in the phase space of the coupled clock system. There are two types of fluxes:
\begin{align}
J_i&=k[e_g+\frac{2\beta}{m}\sum_jE^\prime(\varphi_{ij})]P_s,\\
J_{ij}&=-\frac{2\Omega}{m}[E^\prime(\varphi_{ij})-\frac{\beta}{m}\sum_k(E^\prime(\varphi_{ik})-E^\prime(\varphi_{jk}))]P_s,
\end{align}
where $J_i$ is the processive flux for the $i$-th clock; $J_{ij}$ is the exchange flux between clock-$i$ and clock-$j$.
Both fluxes are nonzero, which means that continuous energy dissipation is needed to maintain the NESS.
The free energy dissipation rate per oscillator is given by the entropy production rate~\cite{ganhui} (see SI for derivation):
\begin{equation}
\dot{W}=\frac1m\int [\sum_i \frac{J_i^2}{k P_s}+\sum_{i<j}\frac{J_{ij}^2}{\frac{\Omega}{m} P_s}]d\vec{\phi},
\label{power}
\end{equation}
where the two terms in the RHS of Eq.~\ref{power} correspond to the dissipation for phase procession and phase exchange, respectively. %$C_3$ is a three-point function defined as:

\subsection{The energy cost for driving the nonequilibrium transition to synchronization} Following standard convention~\cite{acebron2005}, we define the synchronization order parameter $0\leq r<1$ by
%\begin{equation}
%r\equiv m^{-1}\sum_{j=1}^m \cos \theta_i,
\[
re^{i\psi}\equiv \frac1m\sum_{j=1}^m e^{i \phi_j},
\]
%\label{r_def}
%\end{equation}
%where $\theta_i=\phi_i -m^{-1} \sum_{i=1}^m \phi_i$ is the deviation of the phase of oscillator-$i$ from the average phase.
where $\psi$ is the phase of the collective oscillation. We define the phase fluctuation of oscillator $i$ from that of the mean oscillation as: $\theta_i\equiv \phi_i -\psi$, which
can be described by a distribution $\rho (\theta)$.
In the asynchronous phase, $\rho(\theta)$ is uniform and $r=0$; in the synchronous phase, $\rho(\theta)$ peaks at $\theta=0$ and $r$ becomes finite $(0<r<1)$.

For simplicity, we study a ``ferromagnetic" interaction energy function $E(\phi_i-\phi_j)=-\frac{E_0}{2}\cos(\phi_i-\phi_j)$, with $E_0(>0)$ the coupling strength. By using the exact solution Eq.~\ref{ansatz}, we obtain the steady state distribution for $\rho(\theta)$ in the mean-field limit $m=\infty$ (see SI for simulation results for finite $m$):
%  with $r$ the collective oscillation's amplitude and $\psi$ the collective phase, and $0\leq r\leq 1$. A well(poorly) synchronized system has $r\to 1$($r\to0$).
%In order to analytically track the onset of the collective oscillation, we take the limit of $m\rightarrow\infty$ and propose a mean-field approximation to analyze this onset.
%In this approximation, all the oscillators are supposed to be identically distributed with an one-oscillator density $\rho(\phi,t)$, and we have the order parameter $r\cos\psi=\int_0^{2\pi} \rho (\phi,t)\cos\phi \d\phi$.
%For simplicity here we take a representative specific case, $E(\phi_i-\phi_j)=-\frac{E_0}2\cos(\phi_i-\phi_j)$, then, from \eqref{FP_m}, the equation governing $\rho,\psi$ can be :
%\be
%\frac{\partial\rho}{\partial t}=-\frac{\partial}{\partial\phi}\left(ke_g-\Omega E_0r\sin(\phi-\psi)-\frac\Omega\beta\frac{\partial}{\partial\phi}\right)\rho.
%\label{mean_field}
%\ee
%It's obvious that at steady oscillation state the collective phase $\psi=\psi_0+ke_gt$. By introducing the phase distance to the collective phase, $\theta=\phi-\psi=\phi-\psi_0-ke_gt,$ the steady oscillation solution also has the same form with \eqref{steady_2},
\begin{equation}
\rho (\theta)=Z^{-1}\exp(r\beta E_0\cos\theta).
\label{mft}
\end{equation}
By using the above distribution function $\rho(\theta)$ in the definition for $r$, we obtain the self-consistent equation for the order parameter $r(E_0,\Omega)$ for any given $E_0$ and $\Omega$:
\begin{equation}\label{self-consistence}
r=\int_{0}^{2\pi} \cos\theta \rho(\theta) \d\theta =\frac{I_1(\beta E_0r)}{I_0(\beta E_0r)},
\end{equation}
where $I_0(x)$ and $I_1(x)$ are the modified Bessel functions. %$I_n(x)=\sum_{k=0}^\infty\frac{1}{k!(k+n)!}\left(\frac{x}{2}\right)^{2k+n}.
%$

It can be derived from Eq.~\ref{self-consistence} (see SI for details) that the oscillators are asynchronous, i. e.,  $r=0$ when $\beta E_0<2$. A phase transition to a synchronous state with $r \geq 0$ occurs when $\beta E_0\geq 2$ or equivalently when the exchange frequency $\Omega$ is larger than a critical frequency $\Omega_c(E_0)$:
\begin{equation}
\Omega \ge \Omega_c(E_0) \equiv \frac{2k}{E_0-2}.
\end{equation}

\begin{figure}\centering
%path: see KaiC draft
\includegraphics[width=.9\textwidth]{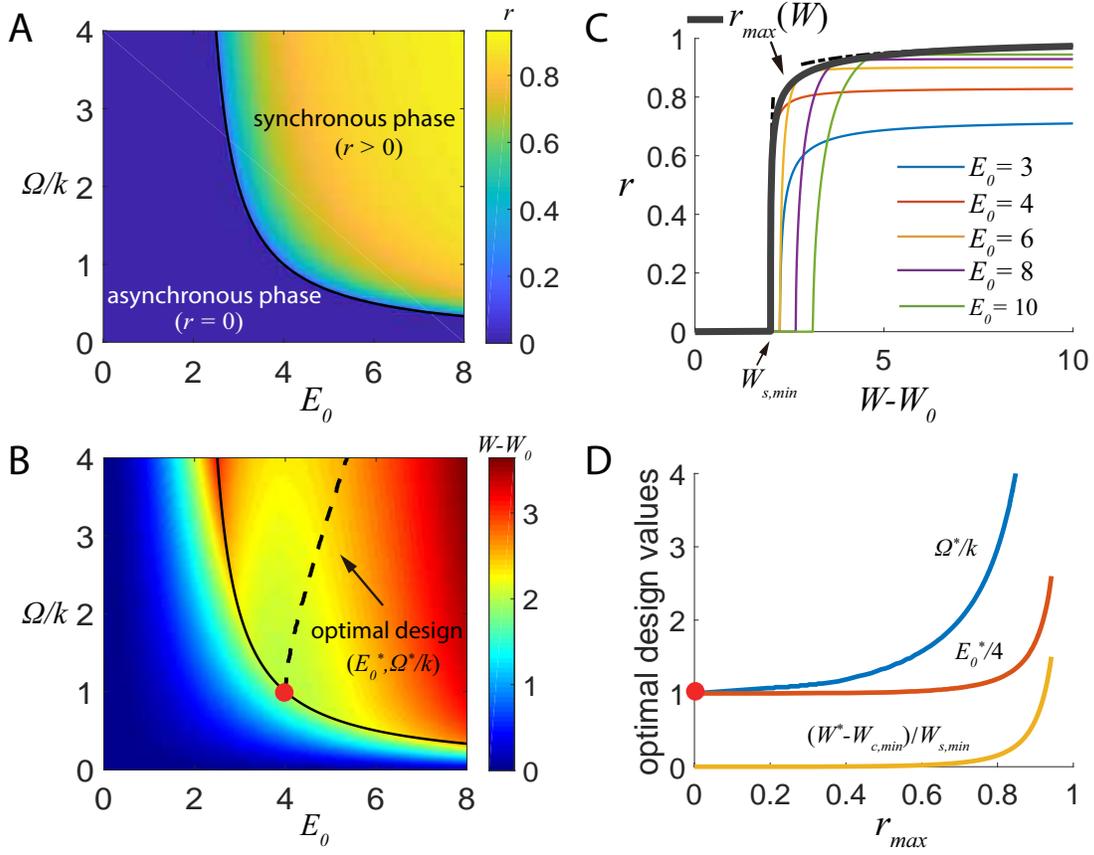}
\caption{Phase diagram and optimal design for synchronization. (A) The synchronization order parameter $r$, and (B) the energy used for the exchange reactions, $W-W_0$, in parameter space $(E_0, \Omega/k)$. The solid line in (A)\&(B) is the phase transition line.  (C) $r$ versus the exchange energy cost $(W-W_0)$ for different values of $E_0$. The thick gray line shows the envelop $r_{max}(W)$, i.e., the maximum $r$ for a given $W$ with its asymptotic behaviors given in Eqs.(\ref{onset_min})\&(\ref{rvsw}) shown by the dotted lines.  (D) The optimal choices $\Omega^*$ and $E_0^*$, and the corresponding energy cost per period $W^*$ to reach the maximum performance $r_{max}$. The optimal design line $(E_0^*,\Omega^*/k)$ is also shown in (B). Parameter $e_g=4\pi$.}
\label{Ksync}
\end{figure}

As shown in the phase diagram Fig.~\ref{Ksync}A, the synchronization transition depends on both the strength and frequency of the exchange reactions. A necessary condition for synchronization is for the exchange energy to be higher than a critical value  $E_0>E_{0,c}\equiv 2$, which is analogous to the critical coupling strength in phase transitions in equilibrium systems such as the Ising model. However, this condition is not enough as synchronization also requires the exchange frequency (rate) to be larger than a critical value $\Omega> \Omega_c (E_0) $. Unlike previously studied cases where nonequilibrium phase transitions are driven by varying temperature \cite{herpich2018collective} or thermal force \cite{nguyen2018phase},
this requirement for kinetic rates studied here is unique to nonequilibrium systems and has no counter part in equilibrium phase transitions.

%{\bf The energy cost of synchronization.}
One hallmark of a nonequilibrium system is that it continuously dissipates energy even in its steady state. But what does it dissipate energy for? Here, we relate the synchronization performance characterized by its order parameter $r$ with the free energy dissipation. By using the phase fluctuation distribution (Eq.~\ref{mft}) in Eq.~\ref{power}, the dissipation rate per oscillator $W=\dot{W}T_p$ in a period $T_p=2\pi/(k e_g)$, can be determined analytically in the limit $m\rightarrow \infty$:
\be\label{diss}
W(E_0,\Omega)=W_0+\frac{2\pi \Omega\beta E_0^2}{ke_g}(\frac{A_2}{2\beta}-A_3),
\ee
where $W_0=2\pi e_g$ is the free energy cost per period for an independent clock, $A_3=\langle\sin(\phi_1-\phi_2)\sin(\phi_1-\phi_3)\rangle = r^2/(\beta E_0)$  and $A_2=\langle\sin^2(\phi_1-\phi_2)\rangle =\frac2{\beta E_0}(1-\frac1{\beta E_0})$ for $\beta E_0\geq2$ are the two- and three-point correlation functions (see SI for derivation). The second term in the RHS of Eq.(\ref{diss}), $W_{ex}(E_0,\Omega)\equiv W(E_0,\Omega)-W_0$, represents the energy cost to power the exchange reactions. The dependence of $W_{ex}$ on $E_0$ and $\Omega$ is shown in Fig.~\ref{Ksync}B.

%In biochemical systems, the interaction strength $E_0$ may be hard to vary, however, kinetic rates can be modulated by enzymes. Thus, we first study the relation between  $r$ and $W$ as we vary the exchange rate $\Omega$ with a fixed  $E_0> 2$.
It is clear from Eq.~\ref{diss} that a finite additional energy cost is needed to increase $\Omega$ to reach the onset of synchronization at $\Omega=\Omega_c=2k/(E_0-2)$. This additional energy cost at the onset of collective oscillation can be defined as the synchronization energy:
\be\label{Esync}
W_{s} \equiv W(\Omega=\Omega_c)-W_0=\frac{ \pi E_0^2}{(E_0-2)e_g}.
\ee
Near the synchronization transition, the order parameter depends on the energy dissipation $W$ in a power-law:
%\begin{equation}\label{onset}
$ r \approx a_w (W-W_c)^{\frac12} $
%\end{equation}
with a mean-field exponent $1/2$ and a constant prefactor $a_w= [2e_g/(\pi E_0)]^{\frac12} (E_0-2)/|E_0-4|$.
The critical energy cost $W_c\equiv W_0+W_{s}$ contains two parts, $W_0$ and $W_{s}$, which are responsible for the oscillation of individual clocks and their synchronization, respectively. %{\color{red} I have decided to use $W_s$ for synchronization energy for simplicity. Please change Fig. 2D and 3C and the SI text accordingly. A small comment for Fig.2C: the label "$r_{max}(W)$" should be consistent with the shade of the envelop line; I also feel that the envelop line is not eye-catchy enough, perhaps make it darker? Finally, Fig. 2C has too much empty space to the left because of the formula for the asymptotic behavior. What if we don't show the formula and make the range of the x-axis to be (0,6) or (0,7)?    } %and synchronization and . which reveals that the onset of synchronized oscillation is driven by dissipation with a power of $1/2$. This result is quite . Making an analogy, we can regard this onset as a second order ``energy driven phase transition".

\subsection{Maximizing synchronization with a fixed energy budget}
Given the dependence of  $r$ and $W$ on $\Omega$ and $E_0$, we next ask what is the maximum achievable synchronization $r_{max}(W)$ for a given energy budget $W$, and what is the optimal design of $E_0$ and $\Omega$ that lead to this maximum performance.

From the dependence of $W_{s}$ on $E_0$ given by Eq. (\ref{Esync}), there exists a minimum synchronization energy $W_{s,min}=8\pi/e_g$ at $E_0=4$ with the corresponding critical exchange frequency equal to the clock frequency $\Omega=2k/(E_0-2)=k$. For $W<W_{c,min} \equiv W_0+W_{s,min}$,  synchronization is {\it impossible}, i.e.,  $r_{max}=0$, for any coupling interaction. For $W\geq W_{c,min}$, $r_{max}\geq 0$, synchronization becomes possible for certain choices of $E_0$ and $\Omega$.

In Fig.~\ref{Ksync}C, the dependence of $r$ on $W$ for different choices of $E_0$ are shown. The (upper) envelop of these $r(W,E_0)$ curves defines $r_{max}(W)$, which is also shown. Near the onset of synchronization $0<W/W_{c,min}-1\ll 1$, $r_{max}$ follows a power law:
\begin{equation}\label{onset_min}
r_{max}(W) \approx c_w (W-W_{c,min})^{\frac{1}{4}}
\end{equation}
with a nontrivial exponent $1/4$ and $c_w=[3e_g/(2\pi)]^{\frac{1}{4}}$. For $W/W_{c,min}\gg 1$, $r_{max}$ approaches $1$ (perfect synchronization) with the difference $(1-r_{max})$  inversely proportional to the energy dissipation (see SI for derivations):
\begin{equation}\label{rvsw}
r_{max}(W)\approx1-\frac{\pi}{e_g(W-W_0)}.
\end{equation}

The optimal choices of $E^*_0 (W)$ and $\Omega^*(W)$ that leads to the optimal performance for a given $W$ are also determined.  In Fig.~\ref{Ksync}D,  we show the optimal exchange interactions ($E_0^*$ and $\Omega^*$) and the corresponding energy cost ($W^*$) versus the achieved maximum synchronization $r_{max}$.  For up to a modestly high level of synchronization $\sim 0.7$, the optimal design for the exchange interaction is to have a roughly constant $E_0$ (slightly higher than $4$) and to tune $\Omega$ higher for higher synchronization. This weak dependence of $r_{max}$ on $E_0^*$ (as long as it is larger than a critical value) is related to the small exponent $1/4$ in Eq.(\ref{onset_min}) (see Methods for a brief discussion and SI for a detailed derivation). This design for efficient synchronization is consistent with biological constraints as the interaction strength $E_0$ may be hard to vary in biochemical systems, but the kinetic rate $\Omega$ can be modulated by enzymes.

\subsection{Synchronization in the Kai system} Our theoretical work is inspired by the Kai system underlying the Cyanobacteria circadian clock. The key molecules in the Kai system are the KaiC proteins that form hexamers under physiological conditions. Each KaiC monomer has two autophosphorylation sites (S-431 and T-432) in its CII domain and the different phosphorylation states of the KaiC hexamer constitute the different phases of the oscillation~\cite{nakajima2005,Rust2007}. The processive transitions between these phosphorylation states (phases) are driven by phosphorylation and dephosphorylation reactions that are controlled by two proteins, KaiA and KaiB, and by transitions between a phosphorylation (P) conformation  and a dephosphorylation (dP) conformation of the hexamer~\cite{Wolde2007,Kondo2007ATP,lin2014,abe2015,chang2015}. A simple model for a single KaiC hexamer is characterized by rates of these reactions as shown in Fig.\ref{Kai}A (see Methods for details of the model). %  for phosphorylation and dephosphorylation -- and powered by ATP hdrolysis.

The molecular mechanism of synchronization in the Kai system is not fully understood. One possibility is the experimentally observed monomer-shuffling phenomenon that allows two KaiC hexamers to exchange monomers when the hexamers are in certain phases of their oscillation~\cite{Kondo2006,emberly2006,Ito2007,Johnson2007,Sasai2007,Sasai2008}, which we focus on in this study. Monomer-shuffling can lead to averaging of phases of the two hexamers involved, which can be described by the phase exchange interaction introduced in our coupled molecular clock model.
%Here we propose a minimal model capturing the essential features of this system, as shown in Fig.\ref{Kai}A. KaiC hexamer has two conformational states\cite{Wolde2007,Kondo2007ATP,lin2014,abe2015,chang2015}which undergoes either phosphorylation (P state) or dephosphorylation (dP state). The number of phosphorylated sites of a KaiC hexamer can range from 0 to $h_{max}$. We take $h_{max}=6$ in our model simply because the Ser-431 site is critical in determining the phase\cite{nishiwaki2007}. The forward and reverse rate for phosphorylation $H_i^p\to H_{i+1}^p$ is $k_p,\gamma_1k_p$, and $k_{dp},\gamma_2k_{dp}$ for dephosphorylation $H_{i+1}^{dp}\to H_i^{dp}$. The transition between P and dP state only happens with reaction $H_6^p\to H_6^{dp}$ and $H_0^{dp}\to H_0^d$ with forward and reverse rate $g,\gamma_3g$. This phosphorylation-dephosphorylation cycel (PdP cycle) together with the conformational change composes the processive oscillating cycle similar to the Poisson clock.
%Similar to \cite{Sasai2008}, we assume monomer shuffling can happen between any hexamers in the same conformation, as shown in fig.\ref{Kai}A. We further assume that the two interacting hexamers tend to reduce their difference of phosphorylation levels.
Explicitly, for any allowed monomer-shuffling reaction $H_i+H_j\to H_k+H_l$ with $i+j=k+l$, where the subscript ``x" is the phosphorylation level of the hexamer $H_x$, the reaction rate is $R\times p_{ij\to kl}$, where $R$ is the shuffling rate per hexamer and
$
p_{ij\to kl}  \propto \exp[-E_s(|k-l|-|i-j|)]
$
with $E_s(>0)$ a phenomenological energy parameter. %{\color{red} I have changed $E_0$ to $E_s$ to avoid confusion with the previous $E_0$ in the general theory. Please change Fig. 3C and the SI accordingly.}%The reverse rate is simply $Rp_{kl\to ij}$.  %Notice that the shuffling rule does not introduce any additional non-equilibrium effects. In fact, it can be proved that the system will reach a unique equilibrium state if $\gamma_{1,2,3}=1,R=0.$
%The collective oscillation's amplitude can be calculated by the average phosphorylation level $h(t)=\sum_ih_i(t)/m$, with $m$ the total number of hexamers. In simulations, we take $k_p=k_{dp}=g$, and $\gamma_{1,2,3}=\gamma$ for simplicity. Now the system's dynamics are determined by $\gamma$ and $R$, as shown in Fig.\ref{Kai}B. The phase-space is separated into two regions by a boundary line $(\gamma_c,R_c)$, crossing which collective oscillation emerges.
We study the effect of monomer shuffling by varying the monomer shuffling rate $R$. In Fig.~\ref{Kai}B, we plot the amplitude (defined as averaged phosphorylation level) of the oscillation versus $R$. It is clear that synchronization, i.e., macroscopic oscillation with a non-zero amplitude appears when the shuffling rate exceeds a critical value $R_c$.

As shown in Fig.~\ref{Kai}C, energy cost increases with the shuffling rate $R$ and the minimum energy cost for synchronization $W_{s}$ (defined the same as in Eq.~\ref{Esync}) depends on $E_s$ and can be bigger than the energy $W_0$ needed for driving oscillation of an individual hexamer.  %oscillation is comparable or bigger than energy needed for driving the  to the processive oscillation dissipation, suggesting the system will consume more energy source, i.e. ATP hydrolysis, to achieve synchronization.
Indeed, an average of $\sim 16$ ATP molecules are hydrolyzed per KaiC monomer during one period~\cite{Kondo2007ATP} while only $2$ ATP molecules per KaiC are needed for the phosphorylation-dephosphorylation clock cycle for the two autophosphorylation sites  in KaiC. What are the additional ATP molecules used for? It is known that they are hydrolyzed by KaiC's  ATPase activity, whose function remains a major mystery in the field.
Here, our theory suggests that the KaiC ATPase activity, powered by the additional ATP molecules, may be responsible for driving synchronization in the Kai system. One immediate consequence is that a reduction in the ATPase activity will suppress any possible energy-consuming synchronization mechanism such as monomer-shuffling\footnote{The other possible synchronization mechanism in the Kai system, i.e., the KaiA differential binding mechanism, also costs energy (details to be published).}  and lead to a reduced synchronization. This prediction should be tested experimentally to help reveal the underlying molecular mechanism for synchronization in the Kai system.  %    %This large amount of ATP consumption is perhaps due to many reasons, and energy cost for synchronization may partially answer for this. %Meanwhile, the critical behavior of the collective amplitude is similar with the second-order phase transition introduced in our theoretical part.

\section{Discussion}
In this paper, we found that coupling interactions such as  between two nonequilibrium noisy clocks violate detailed balance and additional free energy must be spent to maintain synchronization of individual clocks. This is a general result independent of individual clock dynamics and the specific coupling mechanism. The additional energy is used to drive the coupling mechanism to correct the phase error (difference) between noisy clocks. In a simple model where individual clocks interact through exchange reactions, we showed that a finite critical amount of energy dissipation, which depends on both the frequency and the strength of the coupling interaction, is needed to drive the non-equilibrium phase transition from a disordered (asynchronous) state to a ordered (synchronous) state.
We also determined the maximum possible synchronization with a fixed energy budget as well as the optimal design of the exchange interaction for achieving the maximum synchronization efficiently. 

Our theoretical results have important implications for studying biological systems. In particular, the insight on energetics of synchronization makes a previously unsuspected connection between the energy source such as the ATPase activity and the observed synchronization behavior. This connection opens up a new direction to search for possible molecular mechanisms for synchronization in specific systems such as the Kai system, which we are currently pursuing. Finally, our work provides a framework to study thermodynamics of collective behaviors in other extended nonequilibrium systems, such as the flocking dynamics \cite{vicsek1995novel,toner1998flocks,toner1995long}, where global order arises through local interactions between active agents. % and the Turing pattern formation \cite{murray2017self}.   %    This is why synchronization of chemical oscillators requires additional cost and why this cost depends on the noise level. This phenomenon/mechanism may widely exists in biological systems. As we have discussed in Kai system, the biological oscillator may consume additional energy, or additional ATP, to maintain its synchrony and keeps functional. This should not be a unique case; more experiments will prove its generality.
\section{Methods}\label{}
{\bf Derivation of the many-oscillator steady state phase distribution.}
As the interaction energy $E(\phi_i,\phi_j)$ only depends on the phase difference $|\phi_i-\phi_j|$, we would expect the steady state of the system to have rotational invariance, i.e. $P_s(\phi_1+\phi,\phi_2+\phi,...,\phi_m+\phi)=P_s(\phi_1,\phi_2,...,\phi_m)$  for arbitrary $\phi$. Consequently, we have $\sum_i\partial P_s/\partial\phi_i=0$, which could simplify \eqref{FP_m} to:
$
\partial_t P_s=\sum_i\partial_{\phi_i}[
{2\Omega}\sum_{j\neq i}E^\prime(\phi_i-\phi_j)/{m} +(\Omega+k)\partial_{\phi_i}]P_s]=0.
$
The solution is
$
P_s(\phi_1,\phi_2,...,\phi_m)=Z^{-1}\exp(-\beta E_t(\phi_1,\phi_2,...,\phi_m)),
$
with $\beta=\Omega/(\Omega+k), E_t=\frac2m\sum_{i<j}E(\phi_i-\phi_j),$ and $Z$ the normalization constant (partition function).

{\bf The optimal design and its asymptotic behavior.}
For a given energy budget $W^*\geq W_{c,min}$, the maximum possible synchronization $r_{max}(W^*)$ is defined by
$
r_{max}(W^*)\equiv\max_{(E_0,\Omega)\in\{(E_0,\Omega)|W(E_0,\Omega)=W^*\} }r(E_0,\Omega),
$
and the corresponding optimal design values are $(E_0^*,\Omega^*)$.
Considering $r$ increases monotonically with $\Omega E_0/(\Omega+k)$, the optimal values $(E_0^*,\Omega^*)$ are unique. $(E_0^*,\Omega^*)$ can be determined numerically and they are plotted in Fig.~\ref{Ksync}D.

The asymptotic behavior of $r_{max}(W)$ when $W$ is near $W_{c,min}$ and $r_{max}$ is small can be determined as below (see SI for more details). Denoting the small deviations $\delta E=E_0-4$, $\delta\Omega=\Omega-k$ and $\delta W=W-W_{c,min}$, in the limit of $\beta E_0\to2$, we obtain an equation for $r$ combining \eqref{self-consistence}\&\eqref{diss},
%$
%{2\pi}(1-\delta E)r^4/{3e_g}+{\pi}\delta E^2r^2/{2e_g}+{\pi}\delta E^2/{2e_g}-\delta W=0,
%$
from which we solve $r$ as a function of $\delta E$ and $\delta W$  (neglecting higher order terms):
$
r(\delta W,\delta E)=[{3e_g}/{(2\pi)}]^{\frac14}(\delta W^\frac12 +\delta W^{\frac12}\delta E/2-{\pi}\delta W^{-\frac12}\delta E^2/4e_g)^\frac12.
$
For a given $\delta W$, $r$ reaches its maximum when $\delta E=e_g\delta W/\pi.$ Thus we have
$
r_{max}(W)
\approx[{3e_g}/{(2\pi)}]^\frac14(W-W_{c,min})^\frac14
$ as given in Eq.~\ref{onset_min}, and correspondingly 
%\begin{equation}\label{E0*}
$E^*_0=4+\frac{2}{3} r_{max}^4$ 
%\end{equation} 
with the high power $4$ given by the small exponent $\frac{1}{4}$ in Eq.~\ref{onset_min}.  As a result, $E^*_0$ is insensitive to $r_{max}(< 1)$ -- it only increases by $\sim 8\%$ as $r_{max}$ changes from $0$ to $0.7$.
%Substituting into \eqref{self-consistence} in the limit of $r_{max}\to0$, for $\Omega^*$ we have
%$
%\Omega^*=k(1+r_{max}^2).
%$

\begin{figure}\centering
%path: see KaiC draft
\includegraphics[width=.9\textwidth]{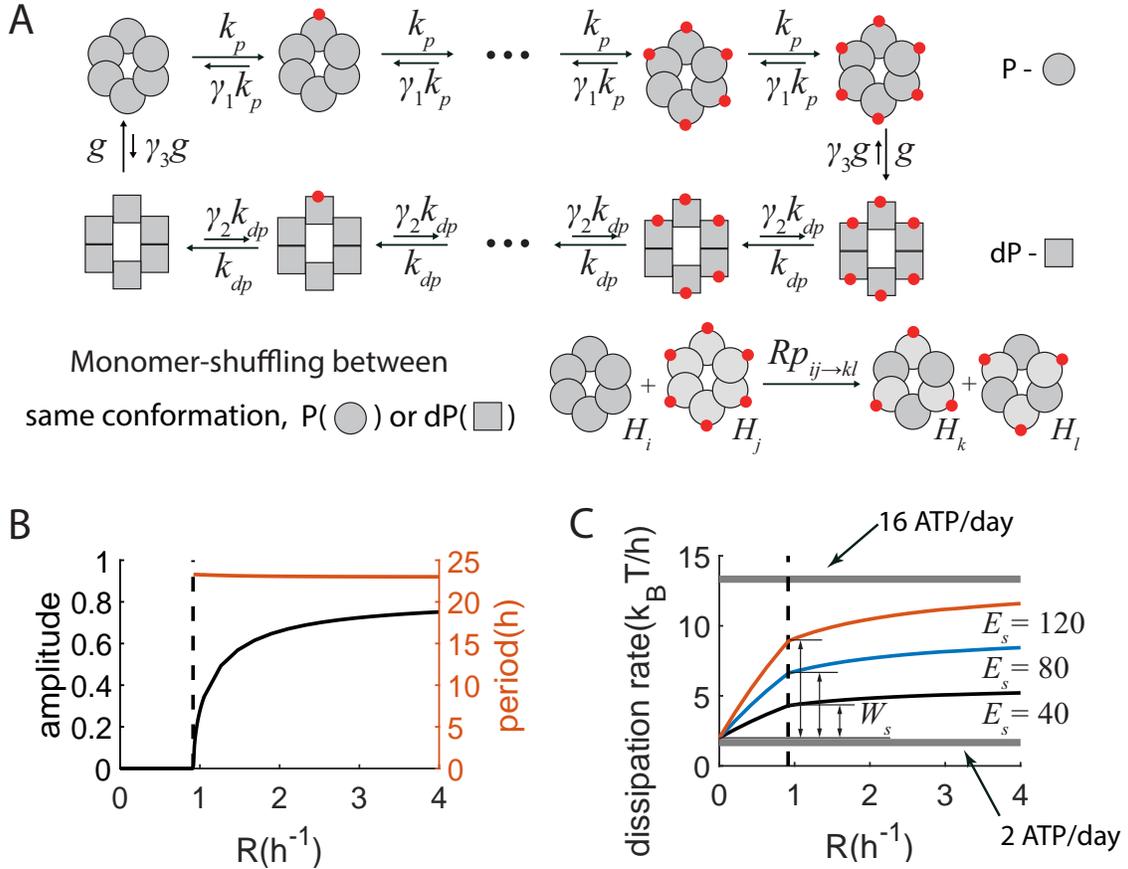}
\caption{The cost of monomer-shuffling for synchronization in the Kai system. (A) Scheme of single hexamer dynamics (top) and monomer shuffling between two hexamers (bottom). The red dot represents the two phosphorylation sites on each KaiC monomer. Shuffling is allowed to happen between two hexamers with the same conformation (P or dP). (B) The amplitude and period of macroscopic oscillation versus the shuffling rate $R$.  A finite critical $R$, labeled by the dotted line (the same as in (C)), is required for the collective oscillation while the period $(\sim 24hr)$ is roughly independent of $R$.  (C) Dissipation rate per Kai monomer versus $R$ for different values of $E_s$. The two gray lines correspond to the minimum energy cost per KaiC monomer for the phosphorylation-dephosphorylation cycle ($2ATP/day$) and the experimentally measured dissipation rate ($\sim 16 ATP/day$), respectively. $1ATP\approx 20 k_B T$ is used here. }
\label{Kai}
\end{figure}

{\bf Details of the model for the Kai system.}
As illustrated in Fig.\ref{Kai}A, there are two kinds of reactions: the processive reactions and monomer shuffling reactions. The processive reactions include phosphorylation, dephosphorylation, and conformational change processes. In our simplified model, a KaiC hexamer has 2 conformations: $P$ and $dP$,and 7 possible phosophorylation states corresponding to the 7 possible numbers (from 0 to 6) of fully phosphorylated KaiC monomers in the hexamer. % in one conformation, with processive reactions between those states representing the two sites phosphorylated/dephosphorylated in one KaiC monomer. 
In its P-conformation, the hexamer favors the phosphorylation reactions with the forward and reverse rates for phosphorylation ($H_i^p\to H_{i+1}^p$) given by $k_p$ and $\gamma_1k_p$, respectively ($\gamma_1<1$). In its dP-conformation, the hexamer favors the dephosphorylation reactions with the forward and reverse rates for dephosphorylation $H_{i+1}^{dp}\to H_i^{dp}$ given by $k_{dp}$ and $\gamma_2k_{dp}$, respectively ($\gamma_2<1$).  The transitions between P and dP conformations only occur with reaction $H_6^p\to H_6^{dp}$ and $H_0^{dp}\to H_0^p$ with forward and reverse rates given by $g$ and $\gamma_3g$, respectively $(\gamma_3<1)$. This phosphorylation-dephosphorylation cycle (PdP cycle) and the conformational change process constitute the (global) processive cycle similar to the Poisson clock shown in Fig.~1A.

Following \cite{Sasai2008}, we assume monomer shuffling happens between hexamers with the same conformation (P or dP). After shuffling, the two hexamers tend to reduce their difference of phosphorylation levels. We explicitly model this process by taking the rate of monomer shuffling reaction $H_i+H_j\to H_k+H_l$ with rate $Rp_{ij\to kl}$, where $R$ is the shuffling rate, and
$
p_{ij\to kl}=Z^{-1}\exp[-E_s(|k-l|-|i-j|)],
$
with $Z=\sum_{k,l}\exp[-E_s(|k-l|-|i-j|)]$ and $E_s$ a phenomenological energy parameter. The reverse rate is simply $Rp_{kl\to ij}$.

Given all these reactions, the concentration of KaiC hexamers in each state (14 states in total) is governed by a set of ordinary differential equations. From simulations of these ODEs, we can compute the amplitude and period of the collective oscillation (Fig.~\ref{Kai}B) as well as the dissipation rate of the whole system (Fig.~\ref{Kai}C). More technical details and parameters used for Fig.~\ref{Kai}B\&C are given in the SI. 

\section{Acknowledgments}
We thank Dr. Thomas Theis for stimulating discussions and critical reading of the manuscript. This work is partially supported by NSFC (11434001,11774011). The work by YT is partially supported by a NIH grant (R01-GM081747).

\section{Data Availability}
All data used to support the findings of this work are available upon request.

\section{Code Availability}
Computer codes used in this work are available upon request.

\bibliographystyle{naturemag}
\bibliography{poisson_osci}
\clearpage

\end{document}

\title{Supplementary Information for ``Nonequilibrium thermodynamics of coupled molecular oscillators: The energy cost and optimal design for synchronization"}
\maketitle
\section{From Master Equation to continuum Fokker-Planck equation}
In this section, we derive the Fokker-Planck equation used in the main text. The general strategy is to write down the master equation, and expand the master equation under small phase separation to obtain the corresponding Fokker-Planck equation.

We consider $m$ identical discrete noisy oscillators labeled by $n_i=1,2,...,N$ and $i=1,2,...,m$. The probability distribution describing this system is $P(n_1,n_2,\cdots,n_{m})$, which is determined by two kinds of reactions: the processive reactions on a full period cycle and the coupling interactions. For each oscillator, a phase variable is introduced $\phi_i=n_i\Delta\phi$ with $\Delta\phi=2\pi/N$.

\subsection{Processive reactions}
The rates of the processive reactions (transitions) are
\begin{gather}
rate(n_i\to n_i+1)=k, \\
rate(n_i+1\to n_i)=\gamma k,
\end{gather}
where $\gamma^N=\Gamma_g$. The master equation is simply:
\be
\frac{dP(n_1,n_2,\cdots,n_{m},t)}{dt}=\sum_i [kP(n_i-1)+\gamma k P(n_i+1)-(1+\gamma)kP(n_i)].
\ee

In the limit of $N\to\infty$, $\Delta\phi\to 0$,  $\Gamma_g$ should keep constant so $\gamma=\Gamma_g^{\frac1N}=\Gamma_g^{\Delta\phi/2\pi}$. In the Fokker-Planck equation, we change to the probability density $P(\phi_1,\phi_2,\cdots,\phi_m)$. Using the phase variables, we have the master equation expanded to second-order:
\begin{align}
\frac{\partial P(\phi_i)}{\partial t}
&=\sum_i [kP(\phi_i-\Delta\phi)+\gamma k P(\phi_i+\Delta\phi)-(1+\gamma)kP(\phi_i)] \notag \\
&\approx\sum_i \left[k \left(-\frac{\partial P}{\partial\phi_i}\Delta\phi+\frac12\frac{\partial^2 P}{\partial\phi_i^2}(\Delta\phi)^2\right)
+\gamma k \left(\frac{\partial P}{\partial\phi_i}\Delta\phi+\frac12\frac{\partial^2 P}{\partial\phi_i^2}(\Delta\phi)^2\right) \right]\notag\\
&=\sum_i\left[-(1-e^{\frac{\Delta\phi}{2\pi}\ln\Gamma_g})k\frac{\partial P}{\partial\phi_i}\Delta\phi+ (1+e^{\frac{\Delta\phi}{2\pi}\ln\Gamma_g})k\frac{\partial^2 P}{\partial\phi_i^2}(\Delta\phi)^2\right]\notag\\
&=\sum_i k(\Delta\phi)^2 \frac{\partial}{\partial\phi_i}(-e_g+\frac{\partial}{\partial\phi_i})P,
\end{align}
where $e_g=-\ln\Gamma_g/2\pi$. The transition rate $k$ is rescaled to $k(\Delta\phi)^2$ as mentioned in the main text.

\subsection{Coupling interactions with different gauges}
The coupling between two oscillators are described by the following exchange reactions:
\begin{gather}
rate[(n_i,n_j)\to(n_i+1,n_j-1)]=r^+(n_i,n_j), \\
rate[(n_i+1,n_j-1)\to(n_i,n_j)]=r^-(n_i+1,n_j-1),
\end{gather}
where the sign $\pm$ means one-step phase increment/decrement for the $i$-th oscillator. These exchange reactions are governed by equilibrium thermodynamics with 
$r^-(n_i+1,n_j-1)/r^+(n_i,n_j)=\exp(-\Delta E_{ij})$ where $\Delta E_{ij}=E(n_i-n_j)-E(n_i-n_j+2)$ is the energy difference between the states $(n_i,n_j)$ and $(n_i+1,n_j-1)$ and $E(n_i-n_j)$ is an even function of the phase difference $n_i-n_j$. By including the coupling reactions, the master equation becomes:
\begin{multline}
\frac{dP(n_1,n_2,\cdots,n_m,t)}{dt}=\sum_i [kP(n_i-1)+\gamma k P(n_i+1)-(1+\gamma)kP(n_i)] \\
+\sum_{i<j}[r^+(n_i-1,n_j+1)P(n_i-1,n_j+1)+r^-(n_i+1,n_j-1)P(n_i+1,n_j-1)
\\
-[r^+(n_i,n_j) + r^-(n_i,n_j)]P(n_i,n_j)],
\end{multline}
where the first line is from the processive motion and the second and third lines are from the coupling reactions.
Using the phase variable in the limit of $\Delta\phi\to 0$, the master equation can be approximated by the Fokker-Planck equation. Below we only express the coupling terms:
\be
\begin{split}
& r^+(\phi_i-\Delta\phi,\phi_j+\Delta\phi)P(\phi_i-\Delta\phi,\phi_j+\Delta\phi)+r^-(\phi_i+\Delta\phi,\phi_j-\Delta\phi)P(\phi_i+\Delta\phi,\phi_j-\Delta\phi) \\
&-(r^+ + r^-)P \\
&\approx\left(\frac{\partial}{\partial\phi_i}-\frac{\partial}{\partial\phi_j}\right)
[(r^--r^+)\Delta\phi P]
+\left(\frac{\partial}{\partial\phi_i}-\frac{\partial}{\partial\phi_j}\right)^2\left(\frac{r^++r^-}{2}(\Delta\phi)^2 P\right).
\end{split}
\ee
Note that $r^+$ is the leaving rate from $(\phi_i,\phi_j)$ to $(\phi_i+\Delta\phi,\phi_j-\Delta\phi)$ and $r^-$ is the arriving rate from $(\phi_i+\Delta\phi,\phi_j-\Delta\phi)$ to $(\phi_i,\phi_j)$.

In the limit $\Delta\phi\to0$, the energy difference becomes $\Delta E_{ij}=-2E^\prime(\phi_i-\phi_j)\Delta\phi$.  However, this energy difference only determines the ratio between the forward and backward exchange reactions. There are many possible choices of how the reaction rates depend on $\Delta E_{ij}$ that preserve this ratio. We call these specific choices of reaction rates different gauges. Below we show that in the continuous phase limit, different gauges lead to the same Fokker-Planck equation.
\begin{enumerate}
\item ``Logistic gauge'',
\begin{align}
r^+(\phi_i,\phi_j)=\frac{\Omega}m\frac{2}{1+\exp(-\Delta E_{ij})},\\
r^-(\phi_i,\phi_j)=\frac{\Omega}m\frac{2}{1+\exp(\Delta E_{ij})}.
\end{align}
In the limit $\Delta\phi\rightarrow 0$, we have
\begin{align}
&r^--r^+\approx\frac{\Omega}m(1-\frac{\Delta E_{ij}}{2})-\frac{\Omega}m(1+\frac{\Delta E_{ij}}{2})=-\frac{\Omega}m\Delta E_{ij}=2E^\prime(\phi_i-\phi_j)\frac{\Omega}m\Delta\phi, \\
&r^-+r^+\approx \frac{2\Omega}m;
\end{align}
\item ``Local gauge" (adopted in the main text),
\begin{align}
r^+(\phi_1,\phi_2)=\frac{\Omega}m\exp(\Delta E_{ij}/2)\\
r^-(\phi_1,\phi_2)=\frac{\Omega}m\exp(-\Delta E_{ij}/2).
\end{align}
Here, we also have
\begin{align}
&r^--r^+\approx\frac{\Omega}m(1-\frac{\Delta E_{ij}}{2})-\frac{\Omega}m(1+\frac{\Delta E_{ij}}{2})=-\frac{\Omega}m\Delta E_{ij}=2E^\prime(\phi_i-\phi_j)\frac{\Omega}m\Delta\phi,\\
&r^-+r^+\approx \frac{2\Omega}m;
\end{align}
\item any gauge that has the form
\begin{equation}
r^\pm=\frac{\Omega}m f(\mp\Delta E_{ij})
\end{equation}
with arbitrary function $f(E)$ satisfies $f(0)= 1$ and $f^\prime(0)=-1/2$.
We still have
\begin{align}
&r^--r^+\approx\frac{\Omega}m(1-\frac{\Delta E_{ij}}{2})-\frac{\Omega}m(1+\frac{\Delta E_{ij}}{2})=-\frac{\Omega}m\Delta E_{ij}=2E^\prime(\phi_i-\phi_j)\frac{\Omega}m\Delta\phi, \\
&r^-+r^+\approx \frac{2\Omega}m;
\end{align}
Notice that the functional form of $f(E)$ is consistent with requirement of equilibrium thermodynamics when $\Delta\phi$ is small,
\begin{equation}
\frac{r^-(\phi_i+\Delta\phi,\phi_j-\Delta\phi)}{r^+(\phi_i,\phi_j)}=\frac{f(\Delta E_{ij})}{f(-\Delta E_{ij})}\approx1+2f^\prime(0)\Delta E_{ij}=1-\Delta E_{ij},
\end{equation}
while thermodynamics requires
\begin{equation}
\frac{r^-(\phi_i+\Delta\phi,\phi_j-\Delta\phi)}{r^+(\phi_i,\phi_j)}=\exp(-\Delta E_{ij})\approx1-\Delta E_{ij}.
\end{equation}
\end{enumerate}

Therefore, for different gauge choices, the same Fokker-Planck equation is reached at the continuum limit
\begin{equation}\label{FPE}
\frac{\partial P}{\partial t}=k\sum_i\frac{\partial}{\partial\phi_i}\left(-e_g+\frac{\partial}{\partial\phi_i}\right)P
+\frac{\Omega}{m}\sum_{i<j}\left(\frac{\partial}{\partial\phi_i}-\frac{\partial}{\partial\phi_j}\right)\left[2E^\prime(\phi_i-\phi_j)+\left(\frac{\partial}{\partial\phi_i}-\frac{\partial}{\partial\phi_j}\right)\right] P,
\end{equation}
where $\Omega$ is rescaled to $\Omega(\Delta\phi)^2$. The steady state distribution is (see Methods section in the main text)
\begin{equation}
P_s(\phi_1,\phi_2,...,\phi_m)=\frac1Z\exp(-\beta E_t(\phi_1,\phi_2,...,\phi_m)),
\end{equation}
with $\beta=\Omega/(\Omega+k), E_t=\frac2m\sum_{i<j}E(\phi_i-\phi_j),$ and $Z$ the normalization constant (partition function).
\section{Mean-field limit}
In the limit of $m\to\infty$, we can apply the mean-field approximation to the system. With this approximation the probability distribution can be decomposed, $P(\phi_1,\phi_2,...,\phi_m,t)=\prod_i\rho(\phi_i,t)$. The Fokker-Planck equation describing $\rho(\phi,t)$ is simply:
\begin{equation}
\frac{\partial\rho}{\partial t}=\frac{\partial}{\partial\phi}\left(-ke_g+\Omega E_0\int\rho(\phi^\prime,t)\sin(\phi-\phi^\prime)\d\phi^\prime+(\Omega+k)\frac{\partial}{\partial\phi}\right)\rho,
\end{equation}
where we used $E(\phi_i-\phi_j)=-\frac{E_0}{2}\cos(\phi_i-\phi_j)$ as in the main text.

To characterize the collective behavior of the system, a complex order parameter is introduced $re^{i\psi}=m^{-1}\sum_je^{i\phi_j}=\langle e^{i\phi}\rangle=\int\rho(\phi,t) e^{i\phi}\d\phi$,
where $0\leq r<1$ is the collective amplitude and $\psi$ is the collective phase.
Using the definition of the order parameter, the coupling term can be written as the interaction with a mean-field, $\int\rho(\phi,t) \sin(\phi-\phi^\prime)\d\phi^\prime=r\sin(\phi-\psi).$ By considering the rotation nature of the collective phase $\psi=\psi_0+ke_g t$ and introducing the phase difference to the collective phase $\theta=\phi-\psi=\phi-\psi_0-ke_gt$, the steady oscillation solution  $\rho_s(\theta)$ should satisfy
\begin{equation}
\frac{\partial\rho_s}{\partial t}=\frac{\partial}{\partial\theta}\left(\Omega E_0r\sin\theta+(\Omega+k)\frac{\partial}{\partial\theta}\right)\rho_s=0,
\end{equation}
from which we can obtain $\rho_s(\theta)$ explicitly,
\begin{equation}\label{ss}
\rho_s(\theta)=\frac1Z\exp(r\beta E_0\cos\theta),
\end{equation}
with $\beta=\Omega/(\Omega+k),Z=\int_0^{2\pi}\exp(r\beta E_0\cos\theta)\d\theta.$
The order parameter $r$ is determined by the self-consistent equation,
\begin{equation}\label{self-consistence}
r=\int_0^{2\pi}\cos\theta\rho_s(\theta)\d\theta=\frac{\int_0^{2\pi}\cos\theta\exp(r\beta E_0\cos\theta)\d\theta}{\int_0^{2\pi}\exp(r\beta E_0\cos\theta)\d\theta}=\frac{I_1(\beta E_0r)}{I_0(\beta E_0r)},
\end{equation}
where $I_0(x)$ and $I_1(x)$ are modified Bessel functions of the first kind,
\[
I_n(x)=i^{-n}J_n(ix)=\sum_{k=0}^\infty\frac{1}{k!(k+n)!}\left(\frac{x}{2}\right)^{2k+n}=\frac1{2\pi}\int_0^{2\pi} e^{x\cos\theta}\cos (n\theta)\d\theta,
\]
for integer $n$.  Denoting $f(x)=I_1(x)/I_0(x)$, it is easy to show that $f^\prime(0)=1/2$ since $I_0(0)=1,I_1(0)=0$. Thus
\eqref{self-consistence} only has one solution $r=0$ when $\beta E_0<2$, which corresponds to an asynchronous state; and has two solutions when $\beta E_0>2$, where the non-zero solution corresponds to the synchronous state. Therefore, $\beta E_0=2$ define the critical line for synchronization in the $(E_0,\Omega)$.

Moreover, we can determine the critical behavior (onset behavior) near $\beta E_0\rightarrow2$ by expanding \eqref{self-consistence} near $r\rightarrow0,$ i.e.
\[
r\approx\frac{\frac{\beta E_0r}{2}+\frac12(\frac{\beta E_0r}{2})^3}{1+(\frac{\beta E_0r}{2})^2},
\]
from which we can determine the critical behavior of $r$ near onset:
\be\label{small_r}
r\approx (\beta E_0 -2)^{\frac12}.
\ee

From \eqref{self-consistence}, we can also find the asymptotic behavior of $r\rightarrow 1$ when $\beta E_0\rightarrow\infty$.
%In this limit, $\theta$ is distributed in a very narrow region near $\cos\theta=1$, which suggests the average can be done in a narrow region  $(-\delta,+\delta)$ , i.e.
%\begin{align}
%r&=\frac1Z\int_{-\pi}^{\pi}\cos\theta\exp(r\beta E_0\cos\theta)\d\theta
%\approx \frac1Z\int_{-\delta}^{\delta}\cos\theta\exp(r\beta E_0\cos\theta)\d\theta \notag \\
%&\approx\frac1Z\int_{-\delta}^{\delta}(1-\frac{\theta^2}{2})\exp(r\beta E_0(1-\frac{\theta^2}{2}))\d\theta
%\approx 1-\frac12\int_{-\infty}^{\infty}\theta^2\cdot\frac1Z\exp(-r\beta E_0\frac{\theta^2}{2}))\d\theta \notag \\
%&=1-\frac{1}{2r\beta E_0}\approx1-\frac{1}{2\beta E_0}.\label{asym}
%\end{align}
The result can be derived from the asymptotic expression of modified Bessel function, i.e. for $x\to\infty$,
\[
I_n(x)\sim\frac{e^x}{\sqrt{2\pi x}}\left(1-\frac{4n^2-1}{8x}+\frac{(4n^2-1)(4n^2-9)}{(8x)^2}+...\right),
\]
thus $I_1(x)/I_0(x)\sim 1 -1/(2x)-1/(8x^2),$ and
\begin{equation}
r=\frac{I_1(\beta E_0r)}{I_0(\beta E_0r)}\sim1-\frac1{2\beta E_0r}-\frac1{8(\beta E_0r)^2},
\end{equation}
which leads to an approximation of $r$ to the order of $(\beta E_0)^{-2}$:
\begin{equation}\label{large_r}
r\approx1-\frac1{2\beta E_0}-\frac3{8(\beta E_0)^2}.
\end{equation}

\section{Energy dissipation and synchronization in the mean-field limit}
\subsection{Energy dissipation in continuum limit}
Energy dissipation rate is given in the discrete system as:
\be
\dot{W}=\sum_i(J_i^+-J_i^-)\ln(J_i^+/J_i^-),
\ee
where $J_i^\pm$ are the forward/backward reaction fluxes and the summation goes over all the \emph{microscopically independent reactions}.
In the continuum limit, the net flux is an infinitesimal flux proportional to $\Delta\phi$, i.e., $J_i^+-J_i^-=\Delta J_i\to0$. This leads to
\begin{equation}\label{exp_diss}
\dot{W}=\sum_i(J_i^+-J_i^-)\ln\left(1+\frac{J_i^+-J_i^-}{J_i^-}\right)\approx\sum_i\frac{(\Delta J_i)^2}{J_i^-}\approx\sum_i\frac{(\Delta J_i)^2}{J_i^+}.
\end{equation}

In our current model, the processive motion on the full period cycle and coupling interactions are independent reactions, so the total dissipation has two terms:
%Generally, the total dissipation rate has the form $\dot{W}_{total}=\sum \dot{W}_i,$
%with the subscript $i$ runs over all \emph{independent processes}.
%The concept of ``independent process" is of great importance. This means flux from different source but via the same process contributes to the same term in entropy production, and the entropy production due to fluxes with similar effect but via different process should be calculate separately.
%Here we start from the original expression of dissipation in discrete case and derive the expression in continuum case by taking the limit $\Delta\phi\to0$.
$\dot{W}=\dot{W}_{cycle}+\dot{W}_{ex}=\sum_i\sum_k\dot{W}_i(\phi_i=k/2m\pi)+\sum_{i<j}\sum_{kl}\dot{W}_{ij}(\phi_i=k/2m\pi,\phi_j=l/2m\pi)$, where
\be
\dot{W}_i(\phi_i)=[kP(\phi_i-\Delta\phi)-\gamma kP(\phi_i)]
\ln\left(\frac{kP(\phi_i-\Delta\phi)}{\gamma kP(\phi_i)}\right),
\ee
and
\begin{multline}
\dot{W}_{ij}(\phi_i,\phi_j)=[r^+(\phi_i-\Delta\phi,\phi_j+\Delta\phi)P(\phi_i-\Delta\phi,\phi_j+\Delta\phi) -r^-(\phi_i,\phi_j)P(\phi_i,\phi_j)] \\
\times\ln\left(\frac{r^+(\phi_i-\Delta\phi,\phi_j+\Delta\phi)P(\phi_i-\Delta\phi,\phi_j+\Delta\phi)
}{r^-(\phi_i,\phi_j)P(\phi_i,\phi_j)}\right).
\end{multline}
In the limit $\Delta\phi\to0$, the summation $\sum_k\to\int\d\phi_i,\sum_{kl}\to\int\d\phi_i\d\phi_j$, and from \eqref{exp_diss} we obtain the expression for $\dot{W}_i$:
\begin{align}
\dot{W}_i(\phi_i)
&\approx\left[k(e_gP-\frac{\partial P}{\partial \phi_i})\Delta\phi\right]^2\cdot\frac1{kP}\notag\\
&=\frac{J_i^2}{kP},\;\; \text{with}\;\; J_i=k\left(e_gP-\frac{\partial P}{\partial \phi_i}\right),
\end{align}
where we have rescaled $k$ to $k(\Delta\phi)^2$. Similarly, we have:
\begin{align}
\dot{W}_{ij}(\phi_i,\phi_j)&\approx\left[(r^+-r^-)P-\left(\frac{\partial (r^+P)}{\partial \phi_i}-\frac{\partial(r^+P)}{\partial \phi_j}\right)\Delta\phi\right]^2\cdot\frac1{r^-P}\notag\\
&\approx\left[-2E^\prime(\phi_i-\phi_j)\frac{\Omega}{m}\Delta\phi\cdot P-\frac{\Omega}{m}\left(\frac{\partial P}{\partial \phi_i}-\frac{\partial P}{\partial \phi_j}\right)\Delta\phi\right]^2\cdot\frac1{\frac{\Omega}{m} P}\notag\\
&=\frac{J_{ij}^2}{\frac{\Omega}{m} P},\;\;\text{with}\;\; J_{ij}=\frac{\Omega}{m}\left[-2E^\prime(\phi_i-\phi_j)P-\left(\frac{\partial P}{\partial \phi_i}-\frac{\partial P}{\partial \phi_j}\right)\right],
\end{align}
where we again rescale $\Omega$ to $\Omega(\Delta\phi)^2.$ Hence the total dissipation rate in the system including all the oscillators in continuum limit is
\begin{equation}\label{diss_con}
\dot{W}_{total}=\int\left[\sum_i\frac{J_i^2}{kP}+\sum_{i<j}\frac{J_{ij}^2}{\frac{\Omega}{m} P}\right]\d\phi_1\d\phi_2...\d\phi_m.
\end{equation}

\subsection{Dissipation in the mean-field limit}
For the specific case $E(\phi_i-\phi_j)=-\frac{E_0}{2}\cos(\phi_i-\phi_j)$ used in our study, the steady state fluxes are
\begin{align}
J_i&=k[e_g+\frac{\beta E_0}{m}\sum_j\sin(\phi_i-\phi_j)]P_s,\\
J_{ij}&=-\frac{\Omega E_0}{m}[\sin(\phi_i-\phi_j)-\frac{\beta}{m}\sum_k(\sin(\phi_i-\phi_j)-\sin(\phi_i-\phi_j))]P_s.
\end{align}
Substituting these fluxes into \eqref{diss_con} and defining the dissipation per oscillator in a period $T_p=2\pi/(ke_g)$ as $W=\dot{W}_{total}T_p/m$, we have in the $m\gg 1$ limit:
%The dissipation rates at large $m$ ($m\gg1$) are:
%\begin{align}
%\dot{W}_{cycle}&=\int\d\vec\phi\sum_i\frac{J_i^2}{kP}
%\sim mk(e_g^2+\frac{4\beta^2}{m^2}(mC_2+m^2C_3))
%\sim m(ke_g^2+4k\beta^2C_3), \\
%\dot{W}_{ex}&=\int\d\vec\phi\sum_{i<j}\frac{J_{ij}^2}{\frac{\Omega}{m} P}
%\sim 2m\Omega
%[C_2-2\frac{\beta}{m}\cdot 2mC_3+\frac{\beta^2}{m^2}(2mC_2+2m^2C_3)]
%\sim 4m\Omega (\frac{C_2}{2}+(\beta^2-2\beta)C_3)],
%\end{align}
%where $C_2=\langle E^{\prime 2}(\phi_1-\phi_2)\rangle,
%C_3=\langle E^\prime(\phi_1-\phi_2)E^\prime(\phi_1-\phi_3)\rangle$. 
%\be\label{diss_origin}
%W=W_0+\frac{8\pi\Omega\beta}{ke_g}(\frac{C_2}{2\beta}-C_3),
%\ee
%where $W_0=2\pi e_g$.
\be\label{diss}
W=W_0+\frac{2\pi\Omega\beta E_0^2}{ke_g}(\frac{A_2}{2\beta}-A_3),
\ee
where $A_2=\langle\sin^2(\phi_1-\phi_2)\rangle,
A_3=\langle\sin(\phi_1-\phi_2)\sin(\phi_1-\phi_3)\rangle$.
In the limit of $m\to\infty$, we can calculate $A_2$ and $A_3$ by using the probability distribution $\rho_s(\theta)$ obtained from the mean-field theory, from which we can relate the collective amplitude $r$ with dissipation rate $W$.

First, we have $\langle\cos\theta\rangle=r$ from definition of $r$.  As $\rho_s(\theta)$ is an even function, it is easy to check that $\langle\sin\theta\rangle=0, \langle\sin\theta\cos\theta\rangle=0$. We also have
\begin{align}
\langle\sin^2\theta\rangle
&=\frac1Z\int\sin^2\theta\exp(r\beta E_0\cos\theta)\d\theta
=-\frac1Z\cdot\frac1{r\beta E_0}\int\sin\theta\d\exp(r\beta E_0\cos\theta) \notag\\
&=\frac1{r\beta E_0}\cdot\frac1Z\int\cos\theta\exp(r\beta E_0\cos\theta)\d\theta
=\frac1{\beta E_0}.
\end{align}
By expanding $A_2$, we have
\begin{align}
A_2&\rightarrow\int\sin^2(\theta_1-\theta_2)\rho_s(\theta_1)\rho_s(\theta_2)\d\theta_1\d\theta_2
\notag\\
&=\frac2{Z^2}\int\sin^2\theta\exp(\beta E_0r\cos\theta)\d\theta\int\cos^2\theta\exp(\beta E_0r\cos\theta)\d\theta \notag\\
&=\left\{
\begin{aligned}
&\frac2{\beta E_0}(1-\frac1{\beta E_0}), &r>0, \\
&\frac12, &r=0,
\end{aligned}\right.
\end{align}
Similarly for $A_3$, we have
\begin{align}
A_3&\rightarrow\int\sin(\theta_1-\theta_2)\sin(\theta_1-\theta_3)\rho_s(\theta_1)\rho_s(\theta_2)\rho_s(\theta_3)\d\theta_1\d\theta_2\d\theta_3
\notag\\
&(\sin^2\theta_1\cos\theta_2\cos\theta_3-\sin\theta_1\cos\theta_1\cos\theta_2\sin\theta_3
-\sin\theta_1\cos\theta_1\sin\theta_2\cos\theta_3+\cos^2\theta_1\sin\theta_2\sin\theta_3)
\notag\\
&=\int\rho_s(\theta_1)\rho_s(\theta_2)\rho_s(\theta_3)\d\theta_1\d\theta_2\d\theta_3\sin^2\theta_1\cos\theta_2\cos\theta_3
\notag\\
&=\frac{r^2}{\beta E_0}.
\end{align}
Substituting $A_2$ and $A_3$ into \eqref{diss}, we have
%we finally have the dissipation per oscillator
\begin{equation}\label{diss_m}
W-W_0
=\frac{2\pi\Omega}{ke_g}(\frac{E_0}{\beta}-\frac1{\beta^2}-E_0r^2).
\end{equation}

\subsection{Synchronization and dissipation near onset}
As stated in the main text, in biochemical systems $E_0$ is hard to vary but rate $\Omega$ can be modulated by enzymes. In the analysis below, we keep $E_0$ fixed so the only free parameter is $\Omega$. At the critical point, $\beta E_0=\Omega_c E_0/(\Omega_c+k)=2$, there is a finite energy for the onset of synchronization:
\begin{equation}
W_{s}\equiv W(\Omega=\Omega_c)-W_0=\frac{\pi E_0^2}{e_g(E_0-2)},
\end{equation}

Near critical point $\beta E_0\to 2$, from \eqref{small_r}, we know that $\beta E_0 - 2=r^2$. Then we have $\Omega=k/[E_0/(2+r^2)-1]$. Substituting into
%\begin{equation}
%\frac1\beta=\frac{E_0}{2+r^2}\approx\frac{E_0}{2}(1-\frac{r^2}{2}),
%\frac1{\beta^2}\approx\frac{E_0^2}{4}(1-r^2),
%\Omega=\frac{k}{\beta^{-1}-1}\approx\frac{2k}{E_0-2}\left(1+\frac{E_0}{2(E_0-2)}r^2\right).
%\end{equation}
\eqref{diss_m} and expanding all the terms to the order of $r^2$, we have
\begin{equation}
W-W_0=\frac{\pi E_0^2}{e_g(E_0-2)}\left(1+\frac{(E_0-4)^2}{2(E_0-2)E_0}r^2\right).
\end{equation}
From the above two equations, we know that near the onset of synchronization the order parameter is related to dissipation in a power law with a power of $1/2$ (except for a singular point at $E_0=4$), i.e.
\begin{equation}
r\sim\left[\frac{2e_g(E_0-2)^2}{\pi E_0(E_0-4)^2}(W-W_0-W_{s})\right]^{\frac12}.
\end{equation}
The behavior at $E_0=4$ is discussed in the following section.

\subsection{Maximizing synchronization with a fixed energy budget}
In the mean-field limit, the order parameter $r$ and dissipation $W$ are fully determined from \eqref{self-consistence} and \eqref{diss_m}. A natural question is what is the maximum synchronization $r_{max}$ the system can achieve for a given dissipation $W$. 
In Fig.2C in the main text,  we provide the numerical solution of $r_{max}(W)$, and here we provide details for deriving its asymptotic behaviors in two limits: (1) weak driving limit, i.e. $r_{max}$ is small and $(W-W_0)$ is close to $W_{s,min}$; and (2) strong driving limit, i.e. $r_{max}$ is close to 1 and $W-W_0\gg W_{s,min}$.
\subsubsection{Weak driving limit}
In the weak driving limit, i.e. $E_0\to4, r\to 0,\delta W=W-W_0-W_{s,min}\ll W_{s,min},$ we can expand $\beta E_0$ with $r^2$ from \eqref{self-consistence} to the order of $r^4$,
\[
\frac{\beta E_0}{2}=1+\frac{r^2}{2}+\frac56r^4.
\]
For a fixed $E_0=4+\delta E,\delta E \ll4$, the exchange rate $\Omega=k/[E_0/(2+r^2+5r^4/3)-1]$, and the dissipation rate can also be expanded in the form of
$
\delta W=\delta W_{s}+Ar^2+Cr^4,
$ with $\delta W_{s}=W_{s}-W_{s,min}$ and
\[
A=\frac{\pi E_0(E_0-4)^2}{2e_g(E_0-2)^2}, C=\frac{\pi E_0^2}{e_g(E_0-2)}\left(\frac{5E_0^2-4E_0}{12(E_0-2)^2}-\frac2{E_0-2}-\frac14\right).
\]
Near $E_0=4,$ we know the leading order of these coefficients, $A\approx\frac{\pi}{2e_g}\delta E^2,C\approx\frac{2\pi}{3e_g}(1-\delta E).$ Expanding Eq.(45), we also have $\delta W_{s}\approx\frac{\pi}{2e_g}\delta E^2.$ This relation tells us that $\delta W$ is no smaller than $O(\delta E^2).$ Substituting into the equation of $\delta W$ we have
\begin{equation}
\frac{2\pi}{3e_g}(1-\delta E)r^4+\frac{\pi}{2e_g}\delta E^2r^2+\frac{\pi}{2e_g}\delta E^2-\delta W=0.
\end{equation}
Solving $r^2$ from this equation and neglecting higher order ($o(\delta E^2)$) terms , we finally obtain $r$ as a function of $\delta E$ and $\delta W,$
\begin{equation}
r(\delta W;\delta E)=\left(\frac{3e_g}{2\pi}\right)^{\frac14}\left(\delta W^\frac12 +\frac{\delta W^{\frac12}}{2}\delta E-\frac{\delta W^{-\frac12}}{2}\frac{\pi}{2 e_g}\delta E^2\right)^\frac12.
\end{equation}
Therefore it is easy to see that, for a given $\delta W$, in weak driving limit the optimal choice $E^*_0$ is $E^*_0=4+[e_g(W-W_{c,min})]/\pi$, and 
\begin{equation}
r_{max}=\left(\frac{3e_g}{2\pi}\right)^{\frac14}\left[(W-W_{c,min})^\frac12 +\frac{(W-W_{c,min})^\frac32}{2}\right]^\frac12
\approx\left[\frac{3e_g}{2\pi}(W-W_{c,min})\right]^\frac14,
\end{equation}
which is presented in the main text (Eq.~14). %This power law with a power of $1/4$ is the critical (asymptotic) behavior in the weak driving limit. Actually, this leading term is exactly the critical behavior of $r$ dependence on $W$ for $E_0=4.$
Furthermore, our analysis yield the dependence of the optimal choice $(E^*_0,\Omega^*)$ on $r_{max}$:
\begin{align}
E^*_0&=4+\frac23r_{max}^4, \\
\Omega^*&=k(1+r_{max}^2).
\end{align}
It reveals that, since $(E^*_0-4)$ depends on $r_{max}^4$, for a modestly high level of synchronization, $E^*_0$ is very close to 4 and it is possible to just tune $\Omega$ higher for higher synchronization.

\subsubsection{Strong driving limit}
From our model, we know that increasing either $\Omega$ or $E_0$ can lead to increased synchronization. However, the effect of increasing $\Omega$ saturates because the effective temperature  $T_{eff}=1+k/\Omega $ can only be reduced to $1$ (the thermal temperature) as $\Omega$ increases to $\infty$. After the saturation, synchronization can only be increased further by increasing $E_0$.

%Here we provide a rough estimate of the maximum synchrony for a given dissipation. 
Based on the above argument, we know that the maximum synchronization is achieved in the parameter regime with $\beta E_0\gg 2$ and $\beta$ near 1. In fact, from the asymptotic behavior of $r$ at large $\beta E_0$ (\eqref{large_r}), maximizing $r$ is indeed equivalent to maximizing $\beta E_0$. Thus we can substitute \eqref{large_r} into \eqref{diss_m} and obtain
\[
W-W_0\approx \frac{2\pi E_0}{e_g}-\frac{2\pi}{\beta e_g}+\frac{\pi\Omega}{ke_g\beta^2 E_0},
\]
from which we get
\[
\beta E_0\approx\frac{\beta e_g}{2\pi}(W-W_0)-\frac{\Omega}{2k\beta E_0}.
\]
In the strong driving limit, both $E_0$ and $(W-W_0)$ should both be large. Thus from the above equation $(W-W_0)$ should at least be of the order $O(E_0)$. Denoting $\Omega/k=\epsilon^{-1},\epsilon\ll 1,$ we have $\beta\approx1-\epsilon$ and the above equation can be expanded in terms of $\epsilon$:
\begin{equation}\label{betaE}
\beta E_0\approx\frac{e_g}{2\pi}(W-W_0)-\frac{e_g}{2\pi}(W-W_0)\epsilon-\frac1{2\epsilon E_0}-\frac1{2E_0}.
\end{equation}
For a fixed $E_0$, it is obvious that the right hand side (RHS) of \eqref{betaE} is maximized when $\epsilon=\sqrt{\pi/[e_g(W-W_0)E_0]}$. As a result, it confirms that the first term is the leading term of RHS when optimized. Hence 
\begin{align}
 E^*_0&\approx\frac{e_g}{2\pi}(W-W_0), \\
\Omega^*&\approx\frac{ke_g}{\sqrt2\pi}(W-W_0)=\sqrt2kE_0^*,
\end{align} 
where we only keep the leading terms, and
\begin{equation}
r_{max}\approx1-\frac1{2\beta E_0^*}\approx1-\frac{\pi}{e_g(W-W_0)}.
\end{equation}
%Substituting into \eqref{large_r}, to the leading order (terms of $(W-W_0)^{-1}$) we have
%\begin{equation}\label{asy_large}
%r<1-\frac1{2\beta E_0}
%<1-\frac{1}{2\beta[\frac{e_g}{2\pi}(W-W_0)+\frac1\beta]}
%<1-\frac{1}{2[\frac{e_g}{2\pi}(W-W_0)+1]}
%\approx1-\frac{\pi}{ e_g(W-W_0)},
%\end{equation}
%where we used $\beta<1$ and neglected the higher order ($(W-W_0)^{-2}$ or smaller) terms. Hence we know that $r_{max}<1-{\pi}/{[e_g(W-W_0)]}.$
This is the upper bound (asymptotically) for the maximum synchronization that can be achieved with a given dissipation in the strong driving limit. In addition, the optimal choice  $(E_0^*,\Omega^*)$ is  inversely proportional to $(1-r_{max})$: 
\[
E_0^*=\frac1{2(1-r_{max})}, \Omega^*=\frac{\sqrt2k}{2(1-r_{max})}.
\]

\section{Simulations of finite systems}
When there is a finite number $(m)$ of oscillators, we can study the synchronization dynamics by direct simulations of the Langevin equations:
\be
\frac{\d\phi_i}{dt}=ke_g+\eta_i(t)+\sum_{j\neq i}(-2E^\prime(\phi_i-\phi_j)+\xi_{ij}(t)),
\ee
where $\eta_i$ is the noise for the processive reactions in the $i$-th oscillator and $\xi_{ij}$ is the noise in the exchange reactions between oscillators $i$ and $j$. The correlations of these noises are given by:   
\begin{align}
\langle\eta_i(t)\eta_j(t^\prime)\rangle &= 2k\delta_{ij}\delta(t-t^\prime),     \\
\langle\xi_{ij}(t)\xi_{kl}(t^\prime)\rangle&=\frac{2\Omega}{m}(\delta_{ik}\delta_{jl}-\delta_{il}\delta_{jk})\delta(t-t^\prime).
\end{align}
It is easy to show that the Langevin equations above give the same Fokker-Planck equation as \eqref{FPE}. We choose $E(\phi_i-\phi_j)=-\frac{E_0}{2}\cos(\phi_i-\phi_j)$ in our simulations as in the main text. 

We calculate the averaged order parameter $r$ by averaging the instantaneous order parameter $r_t$ over a long time $T\gg T_p$, i.e. $r_s=T^{-1}\int_0^Tr_t\d t.$ The comparison between the finite-m cases and the mean-field limit is shown in Fig.~S1. Due to stochastic fluctuations, $r_s$ is finite for $\beta E_0\le 2$. However, $r_s$ approaches the mean-field limit of $r$ with their difference, i.e., the finite size effect, decreases as $1\sqrt{m}$ as shown in the inset of Fig.~S1C.% For the finite-m cases, a sharp transition is presented near the critical point $\beta E_0=2$, and when $\beta E_0\gg 2$, they show same asymptotic behavior as the mean-field limit. By increasing $m$, the curves are approaching to the mean-field limit. We would expect when $m\to\infty$, the results will collapse to mean-field results.

%\bibliography{poisson_osci}

\section{Details of simulations of the Kai system model}
All the reactions in the simple model for the Kai system, as shown in Fig.~3A, are described in the main text. Here, we describe the mathematical details of how we solve the model. For a given set of parameters such as those given in Table 1, the concentration of hexamers in each state (denoted by $h_i^{p/dp}$, corresponding to the concentration in $H_i^{p/dp}$ state) can be described by a set of ordinary differential equations:
\begin{align}
\frac{\d h_i^p}{\d t}&=
(kh_{i-1}^p-(1+\gamma_1)kh_{i}^p+\gamma_1kh_{i+1}^p) \notag \\
&+\sum_{j,k,l}(Rp_{kl\to ij}h_k^ph_l^p-Rp_{ij\to kl}h_i^ph_j^p) \notag \\
&+\delta_{i0}(gh_0^{dp}-\gamma_3gh_0^{p})
-\delta_{i6}(gh_6^{p}-\gamma_3gh_6^{dp}),
\end{align}
and
\begin{align}
\frac{\d h_i^{dp}}{\d t}&=
(kh_{i+1}^{dp}-(1+\gamma_2)h_{i}^{dp}+\gamma_2kh_{i-1}^{dp}) \notag \\
&+\sum_{j,k,l}(Rp_{kl\to ij}h_k^{dp}h_l^{dp}-Rp_{ij\to kl}h_i^{dp}h_j^{dp}) \notag \\
&-\delta_{i0}(gh_0^{dp}-\gamma_3gh_0^{p})
+\delta_{i6}(gh_6^{p}-\gamma_3gh_6^{dp}).
\end{align}
%with $h_i$ the concentration of hexamers at state $H_i$. 
With the trajectories from simulation, we can obtain the collective oscillation's amplitude by the average phosphorylation level $h(t)=\sum_ii(h_i^p+h_i^{dp})/(6h_{tot})$ where $h_{tot}$ is the total concentration. The dissipation rate per monomer (plotted in Fig.3C) is $\dot{W}_{system}/(6h_{tot})$, with
\begin{align}
\dot{W}_{system}&=\sum_{i}(kh_{i-1}^p-\gamma_1kh_{i}^p)\ln\left(\frac{kh_{i-1}^p}{\gamma_1kh_{i}^{dp}}\right)
+\sum_{i}(kh_{i-1}^{dp}-\gamma_2kh_{i}^{dp})\ln\left(\frac{kh_{i-1}^{dp}}{\gamma_2kh_{i}^{dp}}\right) \notag \\
&+\frac12\sum_{i,j,k,l}(Rp_{kl\to ij}h_k^ph_l^p-Rp_{ij\to kl}h_i^ph_j^p)
\ln\left(\frac{Rp_{kl\to ij}h_k^ph_l^p}{Rp_{ij\to kl}h_i^ph_j^p}\right) \notag \\
&+\frac12\sum_{i,j,k,l}(Rp_{kl\to ij}h_k^{dp}h_l^{dp}-Rp_{ij\to kl}h_i^{dp}h_j^{dp})
\ln\left(\frac{Rp_{kl\to ij}h_k^{dp}h_l^{dp}}{Rp_{ij\to kl}h_i^{dp}h_j^{dp}}\right) \notag \\
&+(gh_0^{dp}-\gamma_3gh_0^{p})\ln\left(\frac{gh_0^{dp}}{\gamma_3gh_0^{p}}\right)
+(gh_6^{p}-\gamma_3gh_6^{dp})\ln\left(\frac{gh_6^{p}}{\gamma_3gh_6^{dp}}\right)
\end{align}
in the unit of $k_BT$, where the factor $1/2$ originates from the repeated summation.

\clearpage

\begin{figure}\centering
\includegraphics[width=.9\textwidth]{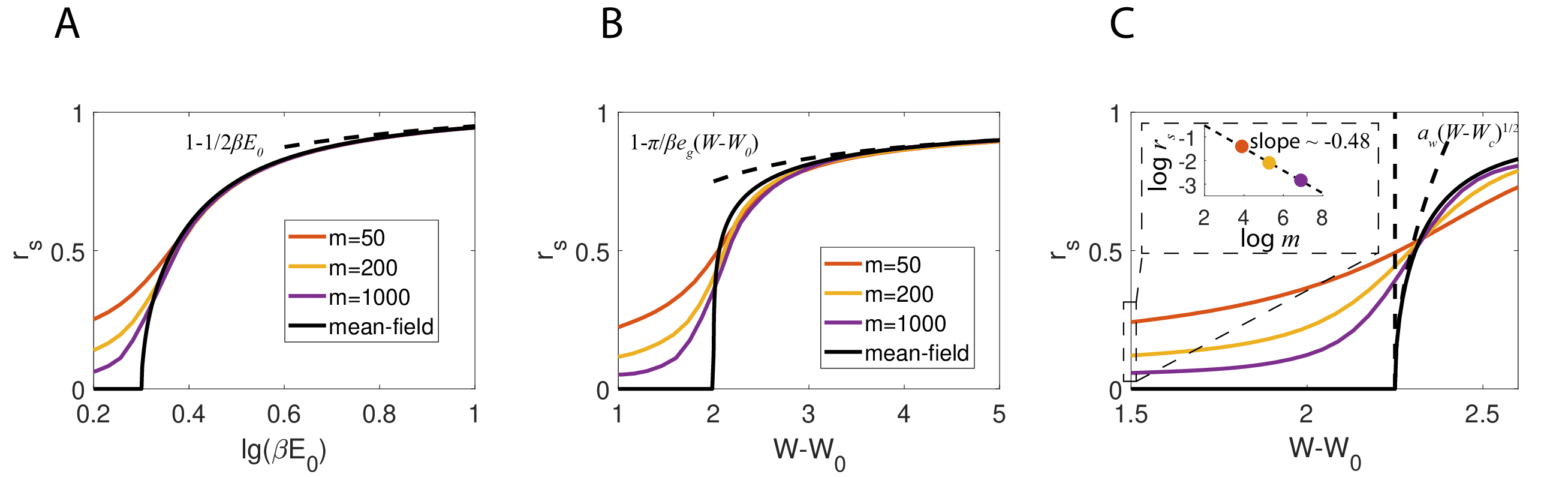}
\caption{Comparing the mean-field results with simulation of finite-m cases with parameters $k=0.5,e_g=4\pi$. (A) Averaged order parameter $r_s$ versus $\beta E_0$. (B) Averaged order parameter $r_s$ versus dissipation by increasing $E_0$. $\Omega=0.5$ is fixed. The black dashed line indicates the asymptotic behavior of $r$ dependent on $W$. (C) Averaged order parameter $r_s$ versus dissipation by increasing $\Omega$. $E_0=6$ is fixed. The dashed box inside shows that for asynchronous state, $W-W_0=1.5,$ $r_s$ converges to zero as $m$ increases approximately with a power law $r_s\sim m^{-1/2},$ indicating the non-zero $r_s$ in asynchronous state is due to stochastic fluctuations. }
\label{finiteM}
\end{figure}

%Parameter tables
\clearpage
\begin{table}
\centering
\begin{tabularx}{0.48\textwidth}{YY}\hline
$k_p (\text{h}^{-1})$ & 0.6 \\
 $k_{dp}(\text{h}^{-1})$ & 0.6 \\
$g(\text{h}^{-1})$ & 0.6\\
$\gamma_{1,2,3}$ & $2\times10^{-9}$ \\
$E_s$(for panel B) & $40$\\
\hline
\end{tabularx}
\caption{Parameters used in simulations for generating results shown in Fig.~3B\&C in the main text. The values of the rates are chosen to have a period to be $\sim 24hr.$ The values of $\gamma_{1,2,3}$ are chosen to have the dissipation per KaiC monomer during a full PdP cycle to be $\sim 2ATP$ because 2 ATP molecules are hydrolyzed for phosphorylation and then dephosphorylation of the two phosphorylation sites (S-431 and T-432) in a KaiC monomer during 24hr. period. The hydrolysis energy of 1 ATP molecule is taken to be $\sim 20 k_B T$ here. }
\end{table}

%In stochastic simulations, we take $k_p=k_{dp}=g$, and $\gamma_{1,2,3}=\gamma$ for simplicity. Now the system's dynamics are determined by $\gamma$ and $R$, as shown in Fig.\ref{Kai}B. The phase-space is separated into two regions by a boundary line $(\gamma_c,R_c)$, crossing which collective oscillation emerges. In Fig.\ref{Kai}C, we plot the amplitude and free energy dissipation by keeping $\gamma$ constant but varying the monomer shuffling rate $R$. It is clear that a non-zero amplitude appears only when the shuffling rate exceeds a critical value. The total dissipation increases as $R$ increases, which is in agreement with our theory. The minimal energy cost $E_{sync}$ (defined same as Eq.\ref{Esync}) to achieve synchronized oscillation is shown in Fig.\ref{Kai}C. Clearly $E_{sync}$ is comparable to the processive oscillation dissipation, suggesting the system will consume more energy source, i.e. ATP hydrolysis, to achieve synchronization. In fact, it has been observed in experiment \cite{Kondo2007ATP} that it requires an amount of 15.8 ATP molecules hydrolysed per KaiC monomer per period, while the basic PdP cycle only requires 2. This large amount of ATP consumption is perhaps due to many reasons, and energy cost for synchronization may partially answer for this.